\documentclass[floatfix,pra,twocolumn,superscriptaddress]{revtex4-2}
\usepackage{amssymb, amsmath}
\usepackage{wasysym}
\usepackage{graphicx}
\usepackage{physics}
\usepackage{units}
\usepackage[american]{babel}

\usepackage[colorlinks=true,urlcolor=black ,citecolor=black,linkcolor=black]{hyperref}

\begin{document}
	\title{Spatially tunable spin interactions in neutral atom arrays}
	
	\author{Lea-Marina Steinert}
	\email[]{lea.steinert@uni-tuebingen.de}
	\affiliation{Max-Planck-Institut für Quantenoptik, 85748 Garching, Germany}
	\affiliation{Munich Center for Quantum Science and Technology (MCQST), 80799 M\"unchen,  Germany}
	\affiliation{Physikalisches Institut, Eberhard Karls Universit\"at T\"ubingen,  72076 T\"ubingen, Germany}
	
	\author{Philip Osterholz}
	\affiliation{Max-Planck-Institut für Quantenoptik, 85748 Garching, Germany}
	\affiliation{Munich Center for Quantum Science and Technology (MCQST), 80799 M\"unchen,  Germany}
	\affiliation{Physikalisches Institut, Eberhard Karls Universit\"at T\"ubingen,  72076 T\"ubingen, Germany}
	
	\author{Robin Eberhard}
	\affiliation{Max-Planck-Institut für Quantenoptik, 85748 Garching, Germany}
	\affiliation{Munich Center for Quantum Science and Technology (MCQST), 80799 M\"unchen,  Germany}
	
	\author{Lorenzo Festa}
	\affiliation{Max-Planck-Institut für Quantenoptik, 85748 Garching, Germany}
	\affiliation{Munich Center for Quantum Science and Technology (MCQST), 80799 M\"unchen,  Germany}
	
	\author{Nikolaus Lorenz}
	\affiliation{Max-Planck-Institut für Quantenoptik, 85748 Garching, Germany}
	\affiliation{Munich Center for Quantum Science and Technology (MCQST), 80799 M\"unchen,  Germany}
	
	\author{Zaijun Chen}
	\affiliation{Max-Planck-Institut für Quantenoptik, 85748 Garching, Germany}
	\affiliation{Munich Center for Quantum Science and Technology (MCQST), 80799 M\"unchen,  Germany}
	
	\author{Arno Trautmann}
	\affiliation{Max-Planck-Institut für Quantenoptik, 85748 Garching, Germany}
	\affiliation{Munich Center for Quantum Science and Technology (MCQST), 80799 M\"unchen,  Germany}
	\affiliation{Physikalisches Institut, Eberhard Karls Universit\"at T\"ubingen,  72076 T\"ubingen, Germany}
	
	\author{Christian Gross}
	\affiliation{Max-Planck-Institut für Quantenoptik, 85748 Garching, Germany}
	\affiliation{Munich Center for Quantum Science and Technology (MCQST), 80799 M\"unchen,  Germany}
	\affiliation{Physikalisches Institut, Eberhard Karls Universit\"at T\"ubingen,  72076 T\"ubingen, Germany}
	
	\date{\today}

	\begin{abstract}
		
		Analog quantum simulations with Rydberg atoms in optical tweezers routinely
		address strongly correlated many-body problems due to the hardware-efficient
		implementation of the Hamiltonian. Yet, their generality is limited, and
		flexible Hamiltonian-design techniques are needed to widen the scope of these
		simulators. Here we report on the realization of spatially tunable interactions
		for XYZ models implemented by two-color near-resonant coupling to Rydberg pair
		states. Our results demonstrate the unique opportunities of Rydberg dressing
		for Hamiltonian design in analog quantum simulators.
		
	\end{abstract}
	
	\maketitle
	
	
	Quantum simulators based on arrays of neutral atoms have proven to be among 
	the most promising platforms to address non-trivial problems of strongly 
	correlated many-body phenomena. This success is based on the optimally 
	hardware-efficient analog implementation of the Hamiltonian under 
	study~\cite{Cirac2012,Georgescu2014, Gross2017, Morgado2021}, which is one of the reasons to
	employ these machines for useful tasks in the so-called NISQ (noisy
	intermediate-scale quantum) era of quantum processors. While such an emulation
	approach eliminates any overhead in control necessities or qubit numbers, it
	strongly restricts the use cases of a specific quantum simulator to problems
	rooted in the device-dependent Hamiltonian. Here, neutral atoms trapped in
	optical tweezer arrays with engineered geometries in two dimensions, laser
	coupled to Rydberg states to induce interactions~\cite{Kaufman2021}, are among
	the most promising platforms~\cite{Bernien2017, DeLeseleuc2019, Madjarov2019,
		Norcia2019, Semeghini2021, Byun2022, Bluvstein2022}. Large system sizes have
	been demonstrated with long coherence times \cite{Madjarov2020, Bluvstein2022},
	enabling the simulation of quantum magnets both in equilibrium
	\cite{Ebadi2021,Scholl2021a} and dynamically~\cite{Bluvstein2021a}.
	
	Rydberg atom-based simulators naturally implement Ising or XY-type Hamiltonians
	with power-law interactions
	\cite{Weimer2010,Zeiher2015,Barredo2015,Browaeys2020}. Recently,
	Floquet-engineered XXZ spin coupling in bulk systems and optical tweezer arrays
	has been demonstrated \cite{Signoles2021,Scholl2021}. Control over the spatial
	interaction profile of Ising systems has also been realized by admixing Rydberg
	character to the ground state, so-called Rydberg dressing \cite{Keating2015,
		Jau2016, gil2014, Zeiher2016, Zeiher2017a, Arias2019, Guardado-Sanchez2021,
		Borish2019}. Lately, a sharply peaked interaction profile has been demonstrated
	by coupling to molecular Rydberg macrodimer potentials \cite{Hollerith2021}.
	One of the biggest remaining challenges is to increase the systems' flexibility
	via universally programmable analog qubit couplings.

	\begin{figure*}[t!]
		\includegraphics[scale=1]{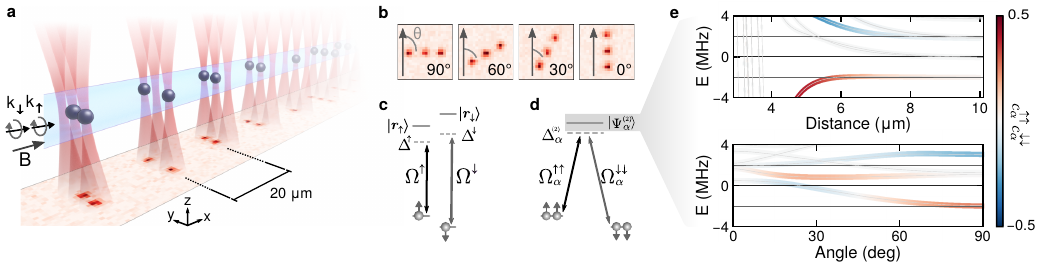}
		
		\caption{\textsf{\textbf{Experimental setup and level schemes.}}
			\textsf{\textbf{a.}}  Illustration of the experimental setting. The Rydberg
			excitation beams (\textbf{k$_\uparrow$} and \textbf{k$_\downarrow$}, light
			blue) are aligned along the magnetic field \textbf{B}, driving $\sigma^-$
			transitions. They illuminate all tweezer groups, each with three linearly
			arranged tweezers (red), which are statistically loaded with atoms (black
			spheres). At the bottom of the illustration, a single shot fluorescence image
			is shown. \textsf{\textbf{b.}} Example of fluorescence images of fully loaded
			tweezer groups for various angles $\theta$ at a distance of $\unit[5.2]{\mu
				m}$. \textsf{\textbf{c.}} On the single-atom level, we couple the electronic
			ground states $\ket*{\uparrow }$ and $\ket*{\downarrow}$ to the Rydberg
			states $\ket*{r_{\uparrow}}$ and $\ket*{r_{\downarrow}}$ with Rabi
			frequencies ($\Omega^{\uparrow}, \Omega^{\downarrow}$) and detunings
			($\Delta^{\uparrow}, \Delta^{\downarrow}$). \textsf{\textbf{d.}} Schematic
			for the flop-flop interaction $J_{ij}^{++}$ between two atoms $i$ and $j$.
			Via adiabatic elimination of the singly excited pair states, we reduce the
			four-photon process to an effective $\Lambda$-scheme. The pairs of ground
			state atoms are coupled with the effective Rabi couplings
			$\Omega_{\alpha}^{\uparrow \uparrow}$ and
			$\Omega_{\alpha}^{\downarrow\downarrow}$ to Rydberg pair states
			$\ket*{\Psi^{(2)}_\alpha}$. $\Delta_{\alpha}^{(2)}$ is the two-photon
			detuning to each $\ket*{\Psi^{(2)}_\alpha}$, which includes interaction
			induced shifts. \textsf{\textbf{e.}} Calculated eigenenergies of
			$\hat{H}_{\text{Ryd}}$ depending on the atom pair distance $d$ at an angle
			of $\theta=\unit[90]{^\circ}$ (upper) and atom pair angle $\theta$ at a
			distance $d=\unit[6]{\mu m}$ (lower). The color scale corresponds to the
			overlap $c_{\alpha}^{\uparrow\uparrow} c_{\alpha}^{\downarrow\downarrow}$.
			The solid lines at $\unit[\pm2]{MHz}$ mark  the energy of the asymptotic
			Rydberg pair state $\ket*{r_{\downarrow}r_{\downarrow}}$ and
			$\ket*{r_{\uparrow}r_{\uparrow}}$. The theoretical results are obtained by
			exact diagonalization of $H_\text{Ryd}$ using the ``pairinteraction''
			software package \cite{Weber2017}. \label{fig:fig1} }
	\end{figure*}

	We report on progress into this direction by the realization of freely
	tunable short-range XYZ-type spin interactions between atoms trapped in
	optical tweezer arrays. The effective \mbox{spin-1/2} system is encoded in two
	electronic ground states and we introduce interactions by two-color
	Rydberg dressing. This allows to engineer the spin-spin couplings in each spin
	direction by the choice of the laser parameters. Our approach uses the
	spatially dependent van-der-Waals (vdW) interactions between different
	\mbox{$m_j$-sublevels} in the Rydberg pair state manifold to design distance
	and angular-dependent couplings of the XYZ Hamiltonian \cite{Glaetzle2015}
	
	\begin{equation}\label{eq:Hamiltonian}
	\hat{H}_\text{XYZ} = \hbar \sum_{i<j}( J^z_{ij} \hat{\sigma}^z_i \hat{\sigma}^z_j
	+ J^{++}_{ij} \hat{\sigma}^+_i \hat{\sigma}^+_j
	+ J^{+-}_{ij} \hat{\sigma}^+_i \hat{\sigma}^-_j ) + h.c.
	\end{equation}
	
\begin{figure}[t!!!]
		\includegraphics[scale=1]{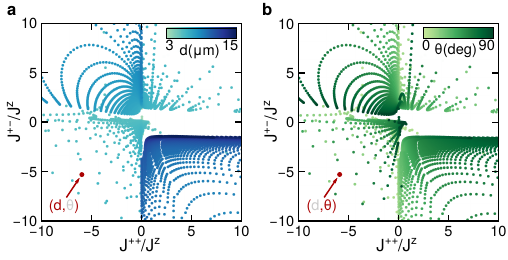}
		\caption{\textsf{\textbf{Tunable XYZ interactions in 1D chains.}} In both panels we show the calculation of the same data points, highlighting the distance (\textsf{\textbf{a.}})
			and angle (\textsf{\textbf{b.}}) dependence of the ratios $J^{++}_{ij}/J^z_{ij}$ and $J^{+-}_{ij}/J^z_{ij}$ for one exemplary set of detunings
			$(\Delta^{\uparrow}, \Delta^{\downarrow}) = 2 \pi \cdot \unit[(1.4,
			-0.6)]{MHz}$. The
			interactions are calculated for distances spaced by $\unit[100]{nm}$ and
			angles in steps of $\unit[1]{^\circ}$. Quadrant II and IV show a smooth
			tunability, while the ratios in quadrant I and III are realized close to Rydberg pair
			state resonances. The latter requires
			higher stability of the control parameters, but may ultimately be beneficial
			for the achievable coherence~\cite{VanBijnen2015}.} \label{fig:fig2}
	\end{figure}

	The Pauli matrices $\hat{\sigma}^z_j$, $\hat{\sigma}^x_j = (\hat{\sigma}^+_j +
	\hat{\sigma}^-_j)$ and $\hat{\sigma}^y_j = i(\hat{\sigma}^-_j -
	\hat{\sigma}^+_j)$ describe a spin-1/2 particle at position~$j$. This
	Hamiltonian distinguishes between three types of spin couplings
	$J^\gamma_{ij}$: The diagonal interaction between dressed ground states $
	J^z_{ij}$, the off-diagonal \grqq flop-flop" $J^{++}_{ij}$, and  \grqq
	flip-flop" $J^{+-}_{ij}$ interactions. While dressing-induced Ising ($J^z$)
	interactions have already been studied in various experiments
	\cite{Zeiher2016, Hollerith2021, Zeiher2017a, Schauss2015, Lienhard2018,
		Borish2019, Kim2010} and programmable long-range interactions have been
	demonstrated in optical cavities~\cite{Periwal2021}, we focus on programmable
	$J^{++}_{ij}$ and $J^{+-}_{ij}$ interactions (see fig.~\ref{fig:fig1}).
	With control over the laser parameters and the geometric arrangement of single
	atoms, we can engineer the relative coupling strength of the spin-spin
	interactions $J^{++}_{ij}/J^z_{ij}$ and $J^{+-}_{ij}/J^z_{ij}$ as visualized
	exemplarily in fig.~\ref{fig:fig2}. We are also able to switch off specific
	couplings globally by the choice of the laser detunings as discussed below. In a
	2D configuration, the situation is even more complex: The angular dependence of the
	interaction provides a unique opportunity to control the nearest-neighbor-
	versus longer-ranged interaction and to realize models featuring  various
	magnetic phenomena, including frustration and topology
	\cite{Glaetzle2015,VanBijnen2015,DeLeseleuc2019,
		Semeghini2021,Wu2022,Tarabunga2022}. Our approach also opens new pathways to
	quantum simulations with practical relevance for the inference of Hamiltonians
	underlying spectra obtained in nuclear magnetic resonance experiments in
	chemistry and biology~\cite{Sels2021}.\\
	
	The physical system we use is an optical tweezer array of single $^{39}$K
	atoms. The spins are encoded in the hyperfine states
	$\ket*{\uparrow}=\ket*{4S_{\nicefrac{1}{2}}\, F=2, m_F = -2}$ and 
	$\ket*{\downarrow}=\ket*{4S_{\nicefrac{1}{2}}F=1, m_F = -1} $. Both states are
	coupled individually to the Rydberg states 
	\mbox{$\ket*{r_{\uparrow}}=\ket*{62P_{\nicefrac{3}{2}},\, m_j =
			-\nicefrac{3}{2}}$} and \mbox{$\ket*{r_{\downarrow}}=\ket*{62P_{\nicefrac{3}{2}},\,
			m_j = -\nicefrac{1}{2}}$} by off-resonant single photon excitation at $286\,$nm
	with the Rabi frequencies $\Omega^{\uparrow}$ and $\Omega^{\downarrow}$, and
	detunings $\Delta^{\uparrow}$ and $\Delta^{\downarrow}$~(see
	Fig.~\ref{fig:fig1}). The choice of beam polarizations suppresses single-atom
	Raman couplings. In this doubly laser-coupled system, rich spin-spin
	interactions emerge, which are rooted in the strong van der Waals (vdW)
	interactions between the addressed Rydberg pair states. For the derivation of
	the spin couplings $J^{++}_{ij}$ and $J^{+-}_{ij}$, we start with diagonalizing
	$\hat{H}_{\text{Ryd}} = \hat{H}_\text{las} + \hat{H}_{\text{int}}$ in the
	Rydberg pair state basis~\cite{Glaetzle2015}. Here, $\hat{H}_\text{las}$ is the
	single atom Hamiltonian in the rotating frame. The vdW Hamiltonian
	$\hat{H}_{\text{int}}$ leads to interactions between the different $m_j$ levels
	in the $62P_{3/2}$ manifold. We admix different components of the vdW pair
	eigenstates to the ground states by laser coupling to obtain the effective interactions
	between the ground states.\\

	The interactions in Eq.~\ref{eq:Hamiltonian} can be understood as four-photon
	processes by adiabatic elimination of all excited
	states~(Supplementary~Information and ref.~\cite{Glaetzle2015}). For example,
	for the flop-flop interactions, the coupling of the $\ket*{\uparrow\uparrow}$
	pair ground state to a Rydberg pair eigenstate $\ket*{\Psi^{(2)}_\alpha}$
	follows by adiabatic elimination of the singly excited state as
	$\Omega_{\alpha}^{\uparrow\uparrow} =(\Omega^{\uparrow})^2 \cdot
	c^{\uparrow\uparrow}_{\alpha}/2 \Delta^{\uparrow}$, where
	$c^{\uparrow\uparrow}_{\alpha}=\braket*{ \Psi^{(2)}_\alpha}{r^\uparrow
		r^\uparrow}$ is the wavefunction overlap of one eigenstate
	$\ket*{\Psi^{(2)}_\alpha}$ in the Rydberg manifold with the asymptotic
	Rydberg pair state~$\ket*{r^\uparrow r^\uparrow}$. The coupling of
	$\ket*{\downarrow \downarrow}$ follows analogously. For sufficiently large
	detuning of the lasers to any coupled state in the Rydberg manifold, we can
	furthermore eliminate the Rydberg pair eigenstates to arrive at an effective
	coupling between ground state atom pairs $i$ and~$j$:
	\begin{equation}
	J^{++}_{ij} = 2\sum_{\alpha}  \frac{\Omega_{\alpha}^{\uparrow \uparrow} \Omega_{\alpha}^{\downarrow \downarrow}}{\Delta_{\alpha}^{(2)}} = \frac{\left(\Omega^{\uparrow} \Omega^{\downarrow} \right)^2}{4\Delta^{\uparrow}\Delta^{\downarrow} }\cdot \frac{ c^{\uparrow\uparrow}_{\alpha}  c^{\downarrow\downarrow}_{\alpha}}{\Delta_{\alpha}^{(2)}} 
	\end{equation}
	
	The Rydberg pair state detuning $\Delta_{\alpha}^{(2)}$ includes vdW
	interaction-induced shifts $U_{\text{vdW},\alpha}$. Spin flips from
	$\ket*{\uparrow \uparrow}$ to $\ket*{\downarrow \downarrow}$ and vice versa
	require a non-zero probability overlap $c^{\uparrow\uparrow}_{\alpha}
	c^{\downarrow\downarrow}_{\alpha}$ provided by the mixing of the
	$m_j$-sublevels. \\
	
\begin{figure}[t!!!]
		\includegraphics[scale=1]{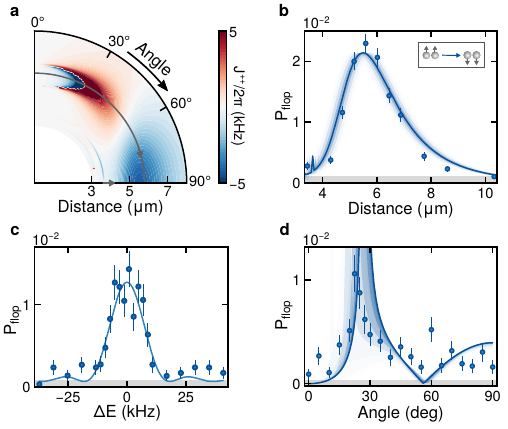}
		\caption{\textsf{\textbf{Flop-flop interactions.}} \textsf{\textbf{a.}}
			Calculation of  $J^{++}$ as a function of $\theta$ and $d$ for
			$\Delta_{\uparrow} =- \Delta_{\downarrow}= 2\pi \cdot \unit[1]{MHz}$. We
			identify a resonance in the spin-spin couplings appearing as a singularity
			around $\theta = \unit[30]{^\circ}$ at a distance of $\unit[5-6]{\mu m}$.
			\textsf{\textbf{b.}} Observed flop-flop probability $P_\text{flop}$ for
			different atom pair distances at $\theta = \unit[90]{^\circ}$. A small false
			positive probability sets the detection limit (grey area) taking into
			account single spin flips and inefficient state preparation and push out.
			Error bars indicate 1 s.e.m. The solid line corresponds to the theoretical
			prediction, where the amplitude has been scaled to match the experimental
			values due to broadening effects. The blue shading indicates the effect
			of the finite radial size of the atomic wavepacket $\sigma_{\text{rad}}$
			~(Supplementary~Information). Each shading represents the interaction
			difference for $\sigma_{\text{rad}}/2$ steps in pair distance in the range of
			$\pm3\sigma_{\text{rad}}$. \textsf{\textbf{c.}}~Flop-flop processes versus
			two-atom Raman detuning $\Delta E = E_{z} - \Delta^\uparrow +
			\Delta^\downarrow$. The fit shows the characteristic $\text{sinc}^2$ envelope
			of a Fourier limited rectangular pulse with a full width half maximum
			$\text{FWHM} = \unit[(18.2 \pm 0.2)]{kHz}$. \textsf{\textbf{d.}}~Angular
			dependence of the flop-flop interaction at a distance of $\unit[5.6]{\mu m}$.
		} \label{fig:fig3}
	\end{figure}

	The derivation of $J^{+-}_{ij}$ starts with two atoms in opposite spin states
	$\ket*{\uparrow\downarrow}$ or $\ket*{\downarrow \uparrow}$. Different from the
	flop-flop interaction case, there are two excitation paths to the Rydberg
	manifold. Via adiabatic elimination of the intermediate singly excited state we
	obtain an effective two-photon coupling $\Omega_{\alpha}^{\uparrow\downarrow}=
	\Omega^\uparrow\Omega^\downarrow \cdot c^{\uparrow \downarrow}_{\alpha} \cdot
	\left(1/4\Delta^\uparrow+  1/4\Delta^\downarrow\right)$. Then, in fourth order
	perturbation theory, we obtain the flip-flop interaction:
	\begin{equation}
	\begin{aligned}
	J^{+-}_{ij} &= 2\sum_{\alpha}  \frac{\Omega_{\alpha}^{\uparrow \downarrow} \Omega_{\alpha}^{\downarrow \uparrow}}{\Delta_{\alpha}^{(2)}} \\
	&= \sum_{\alpha} \frac{\left(\Omega^{\uparrow} \Omega^{\downarrow} \right)^2}{16 \left(\Delta^\uparrow \Delta^\downarrow \right)^2  } \cdot \left(\Delta^\uparrow + \Delta^\downarrow \right)^2 \cdot \frac{ c^{\downarrow \uparrow}_{\alpha} c^{\uparrow \downarrow}_{\alpha}}{\Delta_{\alpha}^{(2)}}
	\end{aligned}
	\end{equation}
	For finite flip-flop interaction, we require a non-zero overlap of $c^{\uparrow
		\downarrow}_{\alpha} c^{\downarrow \uparrow}_{\alpha}$. In the case of
	symmetric detunings $\Delta^\uparrow = - \Delta^\downarrow$, the flip-flop
	interaction is generally canceled by destructive interference of the excitation
	paths. This provides us with sensitive control of $J^{+-}_{ij}$ by choosing the
	excitation laser detuning accordingly. In contrast, energy conservation
	restricts the flop-flop processes and requires the laser detunings to be set to
	$\Delta^{\uparrow}-\Delta^{\downarrow} = E_z$, with $E_z$ the Zeeman splitting
	between $\ket*{r^\uparrow}$ and $\ket*{r^\downarrow}$ (see
	Fig.~\ref{fig:fig3}c).\\

	To study the dependence of the interaction strengths on the geometric
	arrangement experimentally, we select the simplest possible setting of three
	in-line traps with various nearest-neighbor (nn) distances $d$ and angles
	$\theta$ (see Fig.~\ref{fig:fig1}a,b). Here, $\theta$ is the angle between the
	interatomic separation vector $\mathbf{d}$ and the magnetic field $\mathbf{B}$,
	which is set to $\unit[1]{G}$ and defines the quantization axis. We use 14
	replications of this pattern for increased statistics, where the inter-group
	spacing is $\unit[20]{\mu m}$, larger than any interaction range in the system.
	With a first fluorescence image of the atom array, we check for the presence of
	an atom in the trap. We then prepare all atoms in the $\ket*{\uparrow }$ state
	and perform Raman sideband cooling. This allows us to minimize the trap induced
	inhomogeneities by working at the lowest possible tweezer depth of $h\cdot
	\unit[80]{kHz}$ \cite{Lorenz2021}~(Supplementary~Information). We then apply
	two-color Rydberg dressing for $\unit[50]{\mu s}$. Next, we remove all
	$\ket*{\uparrow}$ atoms by a blowout pulse and detect only the remaining atoms
	in the $\ket*{\downarrow}$ state with a second fluorescence image. Comparing
	both fluorescence images allows us to infer the spin interactions by observing
	flipped spins and their correlations.\\
	
\begin{figure}[htp]
		\includegraphics[scale=1]{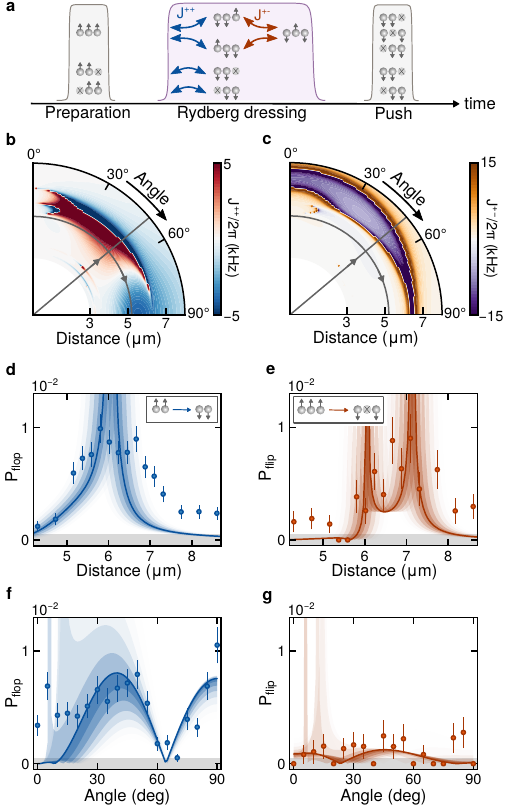}
		\caption{ \textsf{\textbf{Flip-Flop interactions.}} \textsf{\textbf{a.}}
			Experimental sequence. We start with optically pumping all atoms in the
			$\ket*{\uparrow}$ state. This is followed by a period of Rydberg dressing
			and a push out pulse, leaving only atoms in the $\ket*{\downarrow}$ state.
			We post select two initial configurations: The atom in the center plus one
			additional atom or all three atoms are loaded. Depending on this
			configuration, flop-flop (blue arrow) and/or flip-flop processes (orange
			arrow) occur. \textsf{\textbf{b, c.}} Calculation of the distance- and
			angular-dependent flop-flop and flip-flop interactions. \textsf{\textbf{d,
					e.}} Distance dependence of the $J^{++}$ and $J^{+-}$  interactions at an
			angle of $\theta = \unit[50]{^\circ}$ (straight grey line in b, c).
			\textsf{\textbf{f,~g.}}~Angular dependence of the $J^{++}$ and $J^{+-}$
			interactions at a distance of $\unit[5.3]{\mu m}$ (curved grey line
			in~b,~c). Shadings indicate the detection limit and the effects of the
			spatial extension of the atomic wavepacket analogous to figure 2; error
			bars denote 1 s.e.m.}
		\label{fig:fig4}
	\end{figure}

	First, we aim to reveal the induced flop-flop interactions by choosing our
	detuning symmetric $\Delta^\uparrow=-\Delta^\downarrow$, to cancel the
	flip-flop terms. We map the spatial dependence of the interactions by preparing
	the atoms at different distances and angles. We do not observe significant
	single spin flips, confirming the suppression of single-atom Raman processes.
	The $J^{++}$ interaction leads to pairwise spin-flips, which we observe in our
	setting between nearest neighbors. The distance dependence of the pairwise spin
	flips is shown in Fig.~\ref{fig:fig3}, where we scan the atoms' distances at
	$\theta = \unit[90]{^\circ}$ and Rabi couplings of $(\Omega^{\uparrow},
	\Omega^{\downarrow})=2\pi\cdot \unit[(0.52, 0.36)]{MHz}$. The experimental data
	and the amplitude-scaled theoretical expectation are overall in good agreement.
	Differences in theory and experiment emerge from several line-broadening
	effects, such as the finite size of the atoms' thermal wavepacket in radial and
	axial direction in the traps (Supplementary Information), which results in an
	averaging over a range of atom pair separations and angles within the
	radial  ground state wavepacket size of $\sigma^0_\text{rad}=\unit[0.15]{\mu m}$ and
	the axial thermal wavepacket size $\sqrt{2}\sigma^0_\text{ax}\sqrt{k_B T / \hbar
		\omega_\text{ax}}\approx \unit[0.86]{\mu m}$ for the axial trapping frequency
	$\omega_\text{ax} = 2\pi \cdot \unit[1.7]{kHz}$. A second, equally important,
	effect is caused by the line shifts due to tweezer-to-tweezer inhomogeneities
	which result in an average trap depth difference of
	$\overline{\left|\Delta U\right|} = h \cdot \unit[(10.6 \pm 1.6)]{kHz}$. In
	addition, laser phase noise currently limits the dressing time due to a 20-fold
	increased scattering rate appearing as atom loss~\cite{Festa2022}. \\ We
	additionally map out the angular dependence of the flop-flop interaction for a
	fixed distance of $\unit[5.6]{\mu m}$ and Rabi couplings of
	($\Omega^{\uparrow}, \Omega^\downarrow)=2\pi\cdot\unit[(0.55, 0.30)]{MHz}$. In
	this measurement, we cross a singularity in the spin-spin coupling at $\theta
	\approx \unit[30]{^\circ}$ caused by a Rydberg pair state resonance. We
	reproduce a peaked interaction around this resonance, shown in
	Fig.~\ref{fig:fig4}d, while the broadening effects explain the weak atom loss
	by direct Rydberg excitation on resonance~(Supplementary~Information). \\
	
	In the second set of measurements, we switch on both flop-flop and flip-flop
	interactions by setting the detunings to ($\Delta^{\uparrow},
	\Delta^{\downarrow}) = 2 \pi \cdot \unit[(1.4, -0.6)]{MHz}$ and Rabi
	frequencies to ($\Omega^{\uparrow}, \Omega^{\uparrow})=2\pi \cdot \unit[(0.5,
	0.36)]{MHz}$. For the analysis of the spin interactions, we take into account
	different initial states prepared in the statistical loading of the traps
	(see Fig.~\ref{fig:fig4}). These configurations of interest are chosen in
	postselection and correspond to either a fully loaded group
	($\ket*{\uparrow\uparrow\uparrow}$) or groups where only two out of three
	tweezers at the nearest neighbor distance are filled ($\ket*{\uparrow\uparrow
		\circ}$, $\ket*{\circ \uparrow\uparrow}$, where $\circ$ indicates an empty
	site). In the latter configuration, only flop-flop processes occur, while in the case of three atoms also
	flip-flop processes appear. Comparing the two
	fluorescence images before and after the dressing phase allows us to identify
	which interaction processes occurred. More precisely, assuming a
	$\ket*{\uparrow\uparrow\uparrow}$ occupation at the beginning, a flop-flop
	interaction can produce a $\ket*{\uparrow\downarrow\downarrow}$
	($\ket*{\downarrow\downarrow\uparrow}$) arrangement, and flip-flop processes
	introduce the $\ket*{\downarrow\uparrow\downarrow}$ state, after push out
	detected as $\ket*{\downarrow \circ \downarrow}$. The pure flop-flop process
	results in the presence of two atoms at the nearest neighbor distance on the
	second image. We predict a different spatial dependence of $J^{++}$ and
	$J^{+-}$ (see Fig.~\ref{fig:fig4}b,c). Note that the detection method used
	here always requires flop-flop to be present in order to initiate the
	dynamics out of the fully polarized initial state.\\
	Our data reveals the angular- and distance-dependent  $J^{++}_{ij}$
	interactions for an asymmetric detuning in Fig.~\ref{fig:fig4}d,f. In
	addition, we measure a peaked occurrence of the $\ket*{\downarrow \circ
		\downarrow}$ pattern, which we identify as the flip-flop interaction shown in
	Fig.~\ref{fig:fig4}e. This feature reflects the tunability of our system by
	introducing $J^{+-}_{ij}$ interactions for a given laser detuning, atom pair
	distance, and angle. In addition, we probe our system such that the flip-flop
	interaction strength vanishes and only flop-flop interactions occur (see
	Fig.~\ref{fig:fig4}f,g). Here, we scanned the angular dependence of the
	interactions at a fixed distance of $\unit[5.3]{\mu m}$ without crossing a
	Rydberg pair resonance. The minimum in the signal, around
	$\unit[65]{^\circ}$, is caused by interference on the two-atom level.
	Multiple Rydberg pair states $\ket*{\Psi^{(2)}_\alpha}$ with admixtures
	$c^{\uparrow \uparrow}_{\alpha} c^{\downarrow \downarrow}_{\alpha}$ of
	opposite sign contribute such that $J^{++}$ vanishes.\\
	
	In conclusion, we have demonstrated two-color Rydberg-dressing as a new
	technique to achieve tunable, XYZ-type short-range spin interactions in optical
	tweezer arrays. Technical limitations currently prevent us from probing
	coherent interactions (see Supplementary Information). The two leading
	limitations stem from tweezer-to-tweezer line shifts due to array
	inhomogeneities and from laser phase noise. None of these
	are fundamental, in fact, other groups have reported tweezer arrays with less
	than $\unit[1.1]{\%}$ inhomogeneity~\cite{Kim2019}, a factor of 10 improvement
	over our arrays. Laser phase noise can be filtered by optical cavities, as
	demonstrated for Rydberg excitation in ref.~\cite{Levine2018}. Furthermore, it
	is has been shown that the observed Rydberg pair state resonances can be
	utilized to enhance the coherence of Rydberg dressing~\cite{VanBijnen2015}. By
	implementing these measures, we estimate that a maximum figure of merit,
	measured as the product of the peak interaction strength and the coherence
	time, of up to one hundred is reachable with current laser technology.
	This will allow one to realize a flexibly programmable analog quantum
	simulation platform for many-body quantum spin problems. Not only the ratio of
	the spin-interactions in the different channels can be controlled, but also the
	ratio of nearest- to next-nearest-neighbor interactions. This is rooted
	in the non-monotonic spatial dependence of the interaction strength, which can
	also be used to design interactions in two-dimensions for the realization of a
	variety of frustrated geometries~\cite{Glaetzle2015}, static~\cite{Wu2022}
	or dynamic gauge fields~\cite{Tarabunga2022}.\\

	\textbf{Data and materials availability:} The experimental and theoretical
	data and evaluation scripts that support the findings of this study are
	available on Zenodo \cite{Steinert2022}.\\

	\begin{acknowledgments}
		
		\textbf{Acknowledgments:} This project has received funding from the European
		Research Council (ERC) under Grant agreement No. 678580. We also acknowledge
		funding from Deutsche Forschungsgemeinschaft (GR 4741/4-1), within SPP 1929
		GiRyd (GR 4741/5-1), via a Heisenberg professorship (GR 4741/3-1) and  from the
		Alfried Krupp von Bohlen und Halbach foundation.

	\end{acknowledgments}

	\bibliography{bibliography_clean.bib}
	
	\clearpage

	\renewcommand{\thefigure}{S\arabic{figure}}
	\renewcommand{\theequation}{S.\arabic{equation}}
	\setcounter{figure}{0}
	\setcounter{equation}{0}
	
	\section*{Supplementary information}
	This supplementary information document provides information about the experimental sequence and laser setup, a detailed derivation of the interactions, and a discussion of our experimental limitations.
	
	\section{Experimental sequence}
	The optical tweezers are generated using a commercial $\unit[1064]{nm}$ laser aligned onto a liquid crystal spatial light modulator, which imprints a phase pattern onto the beam.
	An in-vacuum mounted objective ($\text{NA}=0.6$) focuses the linearly-polarized beam, obtaining a tweezer array with controlled geometry, of which each trap has a waist of $\unit[0.9]{\mu m}$.
	We load  $^{39}$K atoms into the traps by alternating trapping and cooling light with a frequency of $\unit[1.4]{MHz}$ \cite{Hutzler2017}, one order of magnitude faster than the radial trapping frequency of $\omega_r =2\pi \cdot \unit[158]{kHz}$ ($\omega_z =2\pi \cdot \unit[25]{kHz}$).
	On average, $\unit[50]{\%}$ of the traps are filled with a single atom \cite{Lorenz2021,Festa2022}, and the experimental cycle rate is $\unit[1]{Hz}$.\\
	
	After a first fluorescence image probing the tweezer filling, we optically pump and prepare the atoms in the $\ket*{\uparrow}= \ket*{4S_{\nicefrac{1}{2}} \, F=2, m_F = -2}$ state with $\sigma^-$-polarized pumping and repumping light on the D1-line.
	To quantify the state preparation efficiency, we start by preparing all the atoms in the $\ket*{\uparrow}$ state.
	We then switch on the optical pumping light without the repump light and compare the results for two different polarizations.
	If the light can only drive $\sigma^-$ transitions, the atoms remain in the $\ket*{\uparrow}$ state, which is dark under this illumination on the D1 line.
	Scanning the pulse duration of the pumping light, we fit a $1/e$ optical depumping time $\tau_\text{DP}=\unit[35.93\pm 2.54]{ms}$.
	In contrast, if the polarization is changed such that the light drives $\sigma^+$ or $\pi$ transitions, the atoms are depumped to $F=1$.
	In the latter case, we measure a $1/e$ optical depumping time  $\tau_\text{OP}=\unit[0.20\pm 0.02]{ms}$.
	From this we extract the state preparation efficiency in the $\ket*{\uparrow}$ state to be $P(F=2,m_F=-2) = 1 - \tau_\text{OP}/\tau_\text{DP}=\unit[99.44 \pm 0.08]{\%}$.\\
	
	We then apply Raman sideband cooling (RSC) \cite{Lorenz2021}, which enables us to lower the trap depth to a minimum of $h\cdot \unit[80]{kHz}$ before gravity opens the trap.
	At these low intensities, the light shift induced inhomogeneities of the Rydberg excitation lines are reduced to a few kHz (see section \ref{sec:trapdepths}).
	We continue with a $\unit[50]{\mu s}$ long pulse of both Rydberg dressing lasers (details see section \ref{sec:Rydberg_laser}). The magnetic field strength of $B=\unit[1]{G}$ leads to a Zeeman splitting of the Rydberg states $E_{z} = h \cdot \unit[1.98]{MHz}$.
	
	The Rydberg excitation beams couple the following ground states
	
	\begin{widetext}
		\begin{equation}
		\begin{aligned}
		\ket*{\uparrow }&= \ket*{4S_{1/2}}\ket*{F=2, m_F = -2} = \ket*{4S_{1/2}}\ket*{m_j= -\nicefrac{1}{2}}  \ket*{m_I = -\nicefrac{3}{2}} \\
		\ket*{\downarrow} &= \ket*{4S_{1/2}}\ket*{F=1, m_F = -1} = \ket*{4S_{1/2}}( \ket*{m_j=-\nicefrac{1}{2}}\ket*{m_I = -\nicefrac{1}{2}} - \sqrt{3} \ket*{m_j=\nicefrac{1}{2}}\ket*{m_I = -\nicefrac{3}{2}})/2
		\end{aligned}
		\end{equation}
	\end{widetext}
	to the Rydberg states $\ket*{r_{\uparrow}}=\ket*{62P_{\nicefrac{3}{2}},\, m_j = -\nicefrac{3}{2}}$ and $\ket*{r_{\downarrow}}=\ket*{62P_{\nicefrac{3}{2}},\, m_j = -\nicefrac{1}{2}}$ respectively.
	
	After the Rydberg laser pulse, we remove all remaining atoms in the $F=2$ manifold with a resonant laser pulse on the $\ket*{F=2,m_F=-2}$ to $\ket*{F'=3, m_F'=-3}$ cycling transition of the D2-line.
	We detect the remaining atoms in the $\ket*{\downarrow}$ state with a second fluorescence image.
	Comparing both fluorescence images allows us to deduce the spin interactions based on spin flips and their~correlations.

	\section{Rydberg laser setup \label{sec:Rydberg_laser}}
	
\begin{figure}[t!]
		\includegraphics[width=\columnwidth]{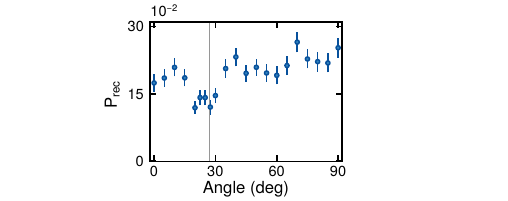}
		\caption{\textsf{\textbf{Atom loss on Rydberg pair resonance. }}
			Measurement settings are the same as in Fig.\,2d in the main text. The gray vertical line marks the predicted position of the pair state resonance. The data shown has been analyzed on a postselected dataset which includes only tweezers with at two nearest neighbor atoms loaded $\ket*{\uparrow \uparrow \circ}$ ($\ket*{\circ \uparrow \uparrow}$). No push out pulse was applied at the end of the sequence to realize a spin spin-insensitive measurement. The atom loss on resonance is observable but weak for our settings.
		}
		\label{fig:fig_si_nopushout}
	\end{figure}

	The Rydberg dressing laser setup consists of a homebuilt ECDL laser at $\unit[1143.5]{nm}$, which is amplified to $\unit[8]{W}$ via a commercial Raman fiber amplifier and then frequency-quadrupled in two consecutive, homebuilt cavity-enhanced doubling stages.
	This results in an output power of $\unit[1]{W}$ at $\unit[286]{nm}$.
	This UV beam is then split into two paths with acousto-optical modulators (AOM) with frequencies of $\pm\unit[230]{MHz}$, which we use for intensity stabilization and bridging the hyperfine ground state splitting.
	The beams are then overlapped and focused onto the atoms, with a horizontal waist of $\unit[40]{\mu m}$ and a vertical waist of $\unit[10]{\mu m}$.
	The Rydberg excitation beams propagate parallel to the magnetic field and drive $\sigma^-$ transitions.\\
	The lifetime of the dressed ground states is proportional to the Rydberg state probability.
	For our parameters and assuming a phase-noise-free laser, we expect a dressed (black-body radiation limited) lifetime of $\tau_{\text{dr}} = \unit[1.7]{ms}$. In contrast, the experimentally observed lifetime is reduced to $\unit{70}{\mu s}$ (for $\Delta_\downarrow = - 2\pi\cdot\unit[0.6]{MHz}$ and $\Omega_\downarrow=2\pi\cdot\unit[0.4]{MHz}$) due to laser noise~\cite{Festa2022}.
	Atom loss due to excitation to Rydberg pair resonances is weak, as shown in Fig.~\ref{fig:fig_si_nopushout}.
	Here, the data has been postselected to only nearest neighbor tweezer pairs initially loaded. In the measurement, we did not apply a push out pulse, to realize spin-insensitive imaging.
	All Rabi frequencies have been measured before the respective set of measurement runs by driving Rabi oscillations without trapping light.
	The typical uncertainty of the Rabi frequency fits is $\unit[0.01]{MHz}$.
	
	\section{Derivation of the effective interactions \label{sec:intderivation}}
	
	\begin{figure*}[t!!!]
		\includegraphics[scale=1]{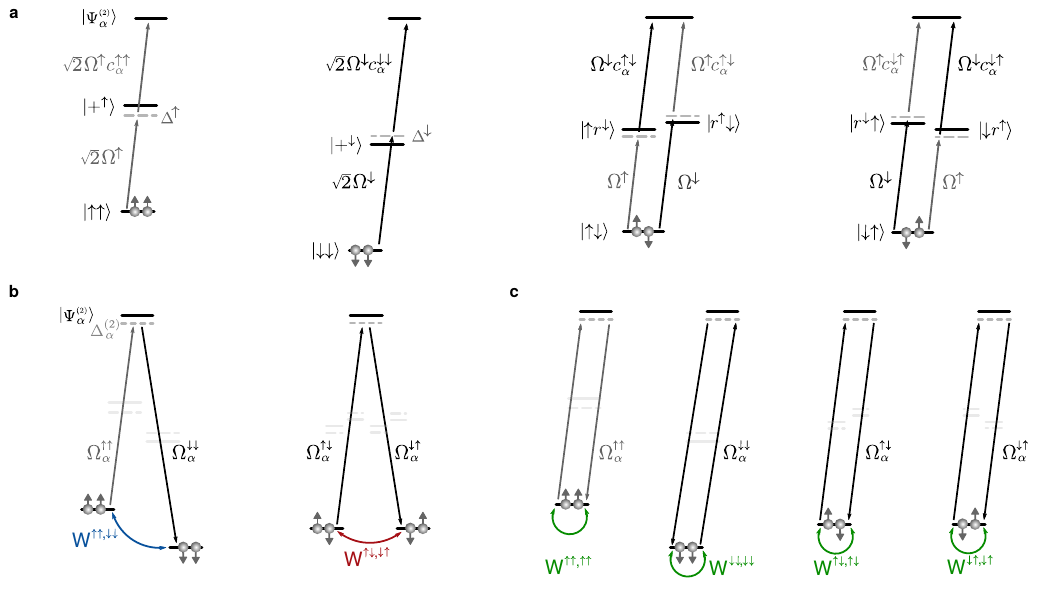}
		\caption{\textsf{\textbf{Stepwise adiabatic elimination.}}
			\textbf{a.} Elimination of singly excited Rydberg states.
			We show the level schemes corresponding to the coupling of different ground state spin-pairs to the Rydberg manifold.
			The singly excited states are adiabatically eliminated to obtain an effective pair state coupling $\Omega_\alpha^{\sigma\sigma'}$ to the eigenstates of the dipolar interaction Hamiltonian $\ket*{\Psi_\alpha^{(2)}}$.
			\textbf{b.} Elimination of doubly excited states.
			The result of the first elimination step is an effective $\Lambda$-system, in which we eliminate the doubly excited states $\ket*{\Psi_\alpha^{(2)}}$ to obtain effective ground state coupling (for unequal initial and final states), or a light shift (for equal initial and final states).
		}
		\label{fig:fig_si_levelscheme}
	\end{figure*}
	
	The effective spin-spin interactions between dressed ground states arise adiabatically by the elimination of the Rydberg levels in the two-atom Hamiltonian $\hat{H}= \hat{H}_\text{las} + \hat{H}_\text{int}$.
	The ground states are laser coupled to the Rydberg states by the laser coupling Hamiltonian 
	
	\begin{equation}
	\begin{aligned}
	\hat{H}_\text{las}/\hbar = \sum_{i=1}^{2} &[\Omega^{\uparrow} (\dyad*{\uparrow}{r^\uparrow}_i + \dyad*{r^\uparrow}{\uparrow}_i)/2  
	+  \Omega^{\downarrow} (\dyad*{\downarrow}{r^\downarrow}_i \\
	&- \dyad*{r^\downarrow}{\downarrow}_i)/2 + \Delta^\uparrow  \dyad*{r^\uparrow}{r^\uparrow}_i - \Delta^\downarrow  \dyad*{r^\downarrow}{r^\downarrow}_i]
	\end{aligned}
	\end{equation}
	The Rabi frequency $\Omega^{\sigma}$ determines the coupling strength between a ground state $\ket*{\sigma}_i$ and a Rydberg state $\ket*{r^\sigma}_i$ of one atom $i$, with $\sigma \in \{\uparrow,\downarrow\}$.
	For the chosen laser polarizations and states, the dipole matrix elements between $\ket{\uparrow}_i$ and $\ket{r^\downarrow}_i$ and vice versa vanish, such that single-atom Raman transitions are absent.
	The magnetic field dependent single atom detunings are described by $\Delta^\sigma$.

	The pair interaction Hamiltonian $\hat{H}_\text{int}$ arises from the dipolar interactions among the Rydberg states.
	We use the ``pairinteraction'' software package to diagonalize the interaction Hamiltonian~\cite{Weber2017} and to obtain the pair-separation $d$ and -angle $\theta$ dependent eigenstates $\ket*{\Psi^{(2)}_\alpha}$ with eigenenergies $E_\alpha(d, \theta)$.
	The eigenstates can be developed in asymptotic pair states $\ket*{r^m r^n}$ as $\ket*{\Psi^{(2)}_\alpha} = \sum_{nm} c_\alpha^{mn}(d, \theta)  \ket*{r^m r^n}$ with the distance- and angle dependend admixture $c_{\alpha}^{mn}(d, \theta)=\braket*{ \Psi^{(2)}_{\alpha}}{r^m r^n}$.
	To improve the readability, we suppress the explicit $d, \theta$-dependency in the rest of the text.
	
	The dipolar interactions between any pair of Rydberg states $\ket*{r^m r^n}$ can be written in the pair basis $\{\ket*{r^m r^m}, \ket*{r^m r^n}, \ket*{r^n r^m}, \ket*{r^n r^n}\}$ in the form
	
	\begin{equation}\label{eq:H_int}
	\hat{H}_\text{int} =
	\begin{pmatrix}
	V^{mm, mm} & 0 & 0 & V^{mm, nn} \\
	0 & V^{mn,mn} & V^{mn, nm} & 0 \\
	0 &V^{nm, mn} & V^{nm, nm} & 0  \\
	V^{nn, mm}  & 0 & 0 & V^{nn, nn}
	\end{pmatrix}.
	\end{equation}
	
	When adiabatically eliminating the Rydberg states (equivalent to 4-th order perturbation theory, c.\,f.~ref.~\cite{Glaetzle2015}), this form of the interactions is transferred to the dressed ground states and the effective Hamiltonian $\hat{H}_\text{eff}$ reads in the $\{ \ket*{\uparrow\uparrow}, \ket*{\uparrow\downarrow}, \ket*{\downarrow\uparrow}, \ket*{\downarrow\downarrow}\}$ basis:
	
	\begin{equation}\label{eq:H_eff}
	\hat{H}_\text{eff} =
	\begin{pmatrix}
	W^{\uparrow\uparrow,\uparrow\uparrow} & 0 & 0 & W^{\uparrow\uparrow,\downarrow\downarrow} \\
	0 & W^{\uparrow\downarrow,\uparrow\downarrow} & W^{\uparrow\downarrow,\downarrow\uparrow} & 0 \\
	0 &W^{\downarrow\uparrow,\uparrow\downarrow} & W^{\downarrow\uparrow,\downarrow\uparrow} & 0  \\
	W^{\downarrow\downarrow,\uparrow\uparrow}  & 0 & 0 & W^{\downarrow\downarrow,\downarrow\downarrow}
	\end{pmatrix}.
	\end{equation}
	
	In the following, we develop an intuitive picture for deriving the different entries of this effective interaction matrix.
	The general idea is a step-wise elimination of the Rydberg levels, starting with singly excited states to obtain a $\Lambda$-system.
	In the second step, we also eliminate the doubly excited states to arrive at an effective Hamiltonian for the dressed ground states.
	This procedure is illustrated in Fig.~\ref{fig:fig_si_levelscheme}.
	Here, we assume that there are only four relevant (i.\,e.~near-resonantly) laser coupled asymptotic pair states, the states $\{\ket*{r^\uparrow r^\uparrow}, \ket*{r^\uparrow r^\downarrow},\ket*{r^\downarrow r^\uparrow},\ket*{r^\downarrow r^\downarrow}\}$.

	\subsection{The diagonal coupling terms}
	
	For the derivation of $W^{\sigma\sigma,\sigma\sigma}$, we start with adiabatic elimination of the single excited state \mbox{$\ket*{+^\sigma}= (\ket*{\sigma r^{\sigma}}+\ket*{r^{\sigma} \sigma})/\sqrt{2}$}. For large atom distances, we obtain the effective two-photon Rabi couplings $\Omega^{\sigma\sigma} = (\Omega^\sigma)^2 / 2\Delta^\sigma$.
	At short distances, the pair potentials in the $m_j$-subspace of the $62P_{\nicefrac{3}{2}}$ manifold interact with each other via dipole-quadrupole interaction, which leads to avoided crossings and mixing of Rydberg states \cite{Urvoy2015}.
	The corresponding admixture $c_{\alpha}^{\sigma\sigma}=\braket*{ \Psi^{(2)}_{\alpha}}{r^\sigma r^\sigma}$ of $\ket*{r^\sigma r^\sigma}$ in close-by interacting pairstates $\ket*{\Psi^{(2)}_{\alpha}}$ reduces the effective Rabi frequencies to $\Omega^{\sigma\sigma}_{\alpha} = \Omega_\text{eff}^{\sigma\sigma}\cdot c_{\alpha}^{\sigma\sigma}$.
	We then adiabatically eliminate the Rydberg pairstates and subtract the asymptotic value of $W^{\sigma\sigma,\sigma\sigma}$ for $d=\infty$ to eliminate a constant offset.
	\begin{equation}
	W^{\sigma\sigma,\sigma\sigma} = \frac{(\Omega^{\sigma})^4}{4(\Delta^\sigma)^2} \sum_{\alpha} \left(\frac{(c_{\alpha}^{\sigma\sigma})^2}{\Delta_{\alpha}^{(2)}} - \frac{1}{2\Delta^{\sigma}}\right),
	\end{equation}
	with the Rydberg pair state detuning \mbox{$\Delta_{\alpha}^{(2)}= 2\Delta^\sigma - E_\alpha$}.\\
	
	The derivation of the $W^{\sigma\bar{\sigma},\sigma\bar{\sigma}}$ (with $\sigma \neq \bar{\sigma}$) is similar: As there are two excitation paths from $\ket*{\sigma \bar{\sigma}}$ to $\ket*{\Psi^{(2)}_{\alpha}}$, the reduced two-photon coupling is \mbox{$\Omega^{\sigma\bar{\sigma}}_{\alpha} =(\Omega^\sigma \Omega^{\bar{\sigma}} / 4 \Delta^\sigma + \Omega^\sigma \Omega^{\bar{\sigma}}/4\Delta^{\bar{\sigma}})\cdot c_{\alpha}^{\sigma\bar{\sigma}}$}. Again we adiabatically eliminate $\ket*{\Psi^{(2)}_{\alpha}}$ and remove a constant offset by subtracting the $d=\infty$ asymptotic value to obtain:
	
	\begin{equation}
	\begin{aligned}
	W^{\sigma\bar{\sigma},\sigma\bar{\sigma}} =&  \left( \frac{1}{(\Delta^{\sigma})^2} + \frac{1}{(\Delta^{\bar{\sigma}})^2} + \frac{1}{\Delta^{\sigma}\Delta^{\bar{\sigma}}} \right)
	\frac{(\Omega^{\sigma})^2(\Omega^{\bar{\sigma}})^2}{16} \\
	&\cdot \sum_{\alpha} \left( \frac{(c_{\alpha}^{\sigma\bar{\sigma}})^2}{\Delta_{\alpha}^{(2)}} - \frac{1}{2\Delta^\sigma} - \frac{1}{2\Delta^{\bar{\sigma}}} \right)
	\end{aligned}
	\end{equation}

	\subsection{The flop-flop off-diagonal terms}
	
	The flop-flop coupling terms $W^{\sigma\sigma , \bar{\sigma}\bar{\sigma}}$ are derived analogously. The effective Rabi frequencies for the flop-flop interactions are $\Omega^{\sigma\sigma}_{\alpha} = \Omega_\text{eff}^{\sigma\sigma}\cdot c_{\alpha}^{\sigma\sigma}$. Via adiabatic elimination of the Rydberg pair states, we obtain the flop-flop coupling term.
	For the off-diagonal terms, offsets at $d=\infty$ are absent since there are two different asymptotic pair state overlaps involved, and one of them must vanish asymptotically. We obtain:
	
	\begin{equation}
	\begin{aligned}
	W^{\sigma\sigma , \bar{\sigma}\bar{\sigma}} &= \sum_{\alpha} \frac{\Omega^{\sigma\sigma}_{\alpha} \Omega^{\bar{\sigma}\bar{\sigma}}_{\alpha}}{\Delta_{\alpha}^{(2)}} \\
	&= \sum_{\alpha} \frac{(\Omega^{\sigma}\Omega^{\bar{\sigma}})^2}{4\Delta^\sigma\Delta^{\bar{\sigma}}} \cdot \frac{c^{\sigma\sigma}_{\alpha}c^{\bar{\sigma}\bar{\sigma}}_{\alpha}}{\Delta^{(2)}_{\alpha}}
	\end{aligned}
	\end{equation}
	
	\subsection{The flip-flop off-diagonal terms}
	
	For the flip-flop term $W^{\sigma\bar{\sigma},\bar{\sigma}\sigma}$ we adiabatically eliminate the single excited states $\ket*{r^{\sigma}\bar{\sigma}}$ and obtain the reduced two-photon Rabi couplings \mbox{$\Omega^{\sigma\bar{\sigma}}_{\alpha} =(\Omega^\sigma \Omega^{\bar{\sigma}} / 4 \Delta^\sigma + \Omega^\sigma \Omega^{\bar{\sigma}}/4\Delta^{\bar{\sigma}})\cdot c_{\alpha}^{\sigma\bar{\sigma}}$}.
	Here, the destructive interference of the two excitation paths for equal magnitude but opposite sign detunings becomes apparent.
	Via adiabatic elimination of the Rydberg manifold, we obtain the flip-flop coupling term:
	
	\begin{equation}
	\begin{aligned}
	W^{\sigma\bar{\sigma},\bar{\sigma}\sigma} &= \sum_{\alpha} \frac{\Omega^{\sigma\bar{\sigma}}_{\alpha} \Omega^{\bar{\sigma}\sigma}_{\alpha}}{\Delta_{\alpha}^{(2)}} \\
	&= \sum_{\alpha} \left(\frac{\Omega^\sigma \Omega^{\bar{\sigma}}}{4\Delta^\sigma} + \frac{\Omega^{\bar{\sigma}} \Omega^\sigma}{4\Delta^{\bar{\sigma}}}\right)^2 \frac{c^{\sigma\bar{\sigma}}_{\alpha}c^{\bar{\sigma}\sigma}_{\alpha}}{\Delta^{(2)}_{\alpha}}
	\end{aligned}
	\end{equation}

	\subsection{Formulation of an effective spin Hamiltonian}
	
	The XYZ-Hamiltonian written in terms of Pauli matrices $\sigma^x,\sigma^y,\sigma^z$ reads
	
	\begin{equation}
	\hat{H}_\text{XYZ} = \hbar \sum_{ij} [J_{ij}^x \sigma^x_i \sigma^x_j + J_{ij}^y \sigma^y_i \sigma^y_j + J_{ij}^z \sigma^z_i \sigma^z_j],
	\end{equation}
	
	or alternatively in raising/lowering form
	
	\begin{equation}\label{eq:ff}
	\begin{aligned}
	\hat{H}_\text{XYZ} = \hbar \sum_{ij} [&J_{ij}^{+-} (\sigma^+_i \sigma^-_j + \sigma^-_i \sigma^+_j)\\
	+ &J_{ij}^{++} (\sigma^+_i \sigma^+_j + \sigma^-_i \sigma^-_j) + J_{ij}^z \sigma^z_i \sigma^z_j],
	\end{aligned}
	\end{equation}
	
	with $\sigma^x_i=(\sigma^-_i + \sigma^+_i)$, $\sigma^y_i=i(\sigma^-_i - \sigma^+_i)$, the flop-flop coupling $J_{ij}^{++} = (J_{ij}^x - J_{ij}^y)$ and the flip-flop coupling $J_{ij}^{+-} = (J_{ij}^x + J_{ij}^y)$.
	
	By expanding Eq.~\ref{eq:ff} in the ground state pair basis and comparing to Eq.~\ref{eq:H_eff} one identifies
	
	\begin{equation}
	\begin{aligned}
	J^z_{ij} &= W^{\uparrow\uparrow,\uparrow\uparrow}(d_{ij}, \theta_{ij}) + W^{\downarrow\downarrow,\downarrow\downarrow}(d_{ij}, \theta_{ij}) \\&\quad\,- 2W^{\uparrow\downarrow,\uparrow\downarrow}(d_{ij}, \theta_{ij})\\
	J^{+-}_{ij} &= 2 W^{\uparrow\downarrow,\downarrow\uparrow}(d_{ij}, \theta_{ij})\\
	J^{++}_{ij} &= 2 W^{\uparrow\uparrow,\downarrow\downarrow}(d_{ij}, \theta_{ij})
	\end{aligned}
	\end{equation}
	
	where we used $W^{\uparrow\downarrow,\uparrow\downarrow} = W^{\downarrow\uparrow,\downarrow\uparrow}$, $W^{\downarrow\uparrow,\uparrow\downarrow}=W^{\uparrow\downarrow,\downarrow\uparrow}$, and $W^{\uparrow\uparrow,\downarrow\downarrow} = W^{\downarrow\downarrow,\uparrow\uparrow}$.
	For clarity, we also restored the pair-separation and -angle dependence here.

	\section{Experimental limitations}
	
	\begin{figure}[t!!!]
		\includegraphics[width=\columnwidth]{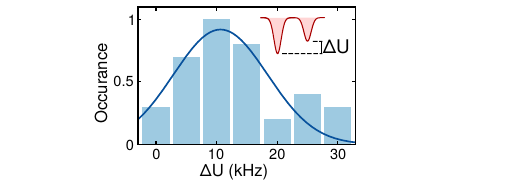}
		\caption{\textsf{\textbf{Tweezer inhomogeneities.}}
			Distribution of the trap depth difference $|\Delta U|$ for two tweezers at the nearest neighbor distance and for the minimum trap depth of $h\cdot \unit[80]{kHz}$. The Gaussian fit (solid line) reveals an average trap depth difference of $\overline{|\Delta U|} = h \cdot \unit[(10.6\pm1.6)]{kHz}$. For our tweezers generated with $\unit[1064]{nm}$ light, the magnitude of the ponderomotive potential for the Rydberg states approximately equals the trap depth for the ground states but is of the opposite sign. Hence, the difference in the line shifts of the ground state-Rydberg transition for nearest-neighbor pairs is about $2|\Delta U|$.
		}
		\label{fig:fig_si_trapdepth}
	\end{figure}

	In our setup, laser noise and inhomogeneous line shifts due to the trapping laser are the main limitations preventing us from probing coherent interactions.
	In the following, we discuss these limitations and their consequences.
	
	\subsection{Laser noise \label{sec:lasernoise}}
	
	Phase noise of our Rydberg excitation laser results in an incoherently enhanced population of the Rydberg states.
	The Rydberg population is determined by $\beta^2=\Omega^2/4 \Delta^2$ for negligible noise, resulting in an excitation rate of $\beta^2 \gamma_r$.
	Here, $\gamma_r^{-1}$ is the Rydberg-state lifetime.
	In our experiment, we measure an about 20-fold increased scattering rate by observing the trap loss~\cite{Festa2022}.
	
	\subsection{Inhomogeneity of trap depths \label{sec:trapdepths}}
	
	To determine the depth of the tweezers, we measure the AC Stark shift on the D1-line by spectroscopy.
	We first prepare the atoms in the $\ket*{F=2, m_F = 2}$ state and set the magnetic field perpendicular to the optical beams such that we probe different polarizations.
	On resonance, the atoms are pumped to the $F=1$ hyperfine manifold.
	We then adiabatically rotate the magnetic field parallel to the direction of the laser beam and remove all atoms in the $F=2$ hyperfine manifold with light resonant to the $\ket*{F=2,m_F=2}$ to $\ket*{F'=3,m_{F'}=3}$ cycling transition of the D2-line.
	We measure the light shift at an average trap depth of $\unit[200]{\mu K}$ and scale the results to the minimal depth used for experiments described in the main text.
	In Fig.~\ref{fig:fig_si_trapdepth} we show the nearest-neighbor tweezer trap depth difference $|\Delta U|$ of a 3x14 tweezer array.

	\subsection{In-trap wavepacket size}
	
	\begin{figure}[t!!!!!]
		\includegraphics[width=\columnwidth]{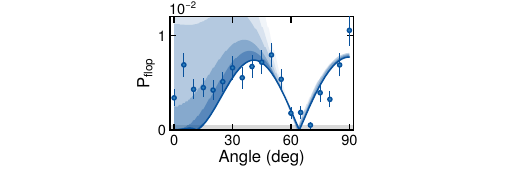}
		\caption{\textsf{\textbf{Influence of the out-of-plane position fluctuations.}}
			Data points are the same as in Fig.\,3f  in the main text. The shading represents the effect of the axial motion of the atoms in traps.
		}
		\label{fig:fig_si_wp_3e}
	\end{figure}
	
	Tweezer inhomogeneities force us to work at a minimal trap depth of $h \cdot \unit[80]{kHz}$.
	This results in a radial (axial) trapping frequency of $\omega_{\text{rad}}=2\pi \cdot \unit[11]{kHz}$ ($\omega_{\text{ax}}=2\pi \cdot \unit[1.7]{kHz}$) with corresponding radial (axial) ground state wavepacket sizes of $\sigma_{\text{rad}}^0=\sqrt{\hbar/(m\omega_{\text{rad}})}=\unit[0.15]{\mu m}$ ($\sigma_{\text{ax}}^0=\unit[0.39]{\mu m}$).
	
	The temperature of our Raman cooled atoms corresponds to $k_B T = h\cdot\unit[4.2]{kHz}$ as measured in ref.~\cite{Lorenz2021}.
	Since the temperature is below the trapping frequency in radial direction we use the ground state wavepacket size to estimate the radial pair-distance fluctuations $\sigma_{\text{rad}} \approx \sqrt{2}  \sigma_{\text{rad}}^0$.
	The factor $\sqrt{2}$ takes the independent motion of the two atoms into account.
	
	The impact of the axial (out-of-plane) wave packet sizes is much weaker and we neglect its effect in the main text.
	Nevertheless, it explains the comparably strong flop-flop interactions for small angles for the measurement shown in Fig.~3f of the main text.
	In Fig.~\ref{fig:fig_si_wp_3e} we show the effect of a thermal wavepacket of size $\sqrt{2}  \sigma_{\text{ax}}^0  \sqrt{k_B T/ \hbar \omega_\text{ax}} \approx \unit[0.86]{\mu m}$  on the flop-flop interactions.
	We use the large temperature limit for the estimation of the position fluctuations here since $k_B T > \hbar \omega_\text{ax}$. 
	The out-of-plane fluctuations result in an averaging over a range of angles, removing the zero in the interactions at a mean angle of $\theta=0^\circ$.

\end{document}